\documentclass{conm-p-l}

\newtheorem{theorem}{Theorem}[section]
\newtheorem{lemma}[theorem]{Lemma}
\newtheorem{proposition}{Proposition}[section]

\theoremstyle{definition}
\newtheorem{definition}[theorem]{Definition}

\theoremstyle{remark}
\newtheorem{remark}[theorem]{Remark}

\numberwithin{equation}{section}

\usepackage{xcolor}

\newcommand{\R}{\mathbb R}
\newcommand{\K}{\mathbb{K}}
\newcommand{\N}{\mathbb N}

\newcommand{\C}{\mathcal{C}}

\newcommand{\F}{\mathcal{F}}
\newcommand{\p}{\mathcal{P}}

\begin{document}

\title{Fr\"olicher structures, diffieties, and a formal KP hierarchy}

\author{Jean-Pierre Magnot}
\address{Univ. Angers, CNRS, LAREMA, SFR MATHSTIC, F-49000 Angers, France
\\ and \\  Lyc\'ee Jeanne d'Arc \\ Avenue de Grande Bretagne, \\ 63000 Clermont-Ferrand, France}
\email{magnot@math.cnrs.fr}

\author{Enrique G. Reyes}
\address{Departamento de Matem\'{a}tica y Ciencia de la Computaci\'{o}n,
	Universidad de Santiago de Chile (USACH), Casilla 307 Correo 2, Santiago,
	Chile}
\email{enrique.reyes@usach.cl ; e\_g\_reyes@yahoo.ca}
\thanks{Research partially supported by the FONDECYT grant \#1201894.}

\author{Vladimir Rubtsov}
\address{Univ. Angers, CNRS, LAREMA, SFR MATHSTIC, F-49000 Angers, France}
\email{volodya@univ-angers.fr}

\subjclass[2020]{ }
\date{ }

\dedicatory{{}{This paper is dedicated to Alexandre M. Vinogradov, 
whose breakthrough insights are at the core of the concepts utilized herein}}

\keywords{Diffieties, Fr\"olicher spaces, Kadomtsev-Petviashvili (KP) hierarchy, Vinogradov sequence}

\begin{abstract}
We propose a definition of a diffiety based on the theory of Fr\"olicher structures. As a consequence, we obtain  a natural Vinogradov sequence and, under the assumption of the existence of a suitable derivation, we can form on it a Kadomtsev-Petviashvili hierarchy which is well-posed.
\end{abstract}

\maketitle

\section{Introduction}
The Erlangen Program of F. Klein and works of S. Lie became a base of the ``Geometric Revolution'' in the 
classical XIX century approach to differential equations (DEs). The next stage of this revolution was manifested 
by applications to DEs of powerful algebro-geometric and algebro-topological methods developed by the French 
School (E. and H. Cartan, Ch. Ehresmann, J. Leray, J.-P. Serre and of course A. Grothendieck). Other 
 very similar or close (differential-geometric) toolbox was developed in Japan by K. Kodaira and M. 
Kuranishi and in US by D. Spencer, H. Goldschmidt, D. Quillen, V. Guillemin and S. Sternberg.

A. M. Vinogradov (whose scientific start and first twenty years of work coincided with the 
so-called ``Golden Age'' of the Moscow Mathematical School) was probably the first person who unified 
successfully two sides of algebro-geometric instrumentaria in his approach to the study of (non-)linear Partial 
Differential Equations ((N)PDEs). He developed and combined ideas of Grothendieck and the differential-geometric   
viewpoint of Ehresmann, and introduced a category whose objects he called {\it diffieties (differential 
varieties)} to study {\it infinitely prolonged differential  equations}. He also proposed a theory for the 
study of diffieties, what has became known as the {\it secondary calculus}, see \cite{Vi}. From a technical 
point of view, the main ingredients of secondary calculus are an appropriate mixture of commutative and 
homological algebra with differential geometry. The concept of a diffiety plays the same role in the theory of 
PDEs that affine algebraic varieties do in the theory of algebraic equations. Various natural characteristics of 
a diffiety and, consequently, of the corresponding system of PDEs, are expressed in terms of this calculus and 
vice versa.

A.M. Vinogradov himself says the following  about the appearance of the notion of a diffiety 
(see \cite{Vin2013}):
\begin{quotation}
... it was understood that various natural differential operators and constructions that are necessary for the 
study of a system of PDEs of order $k$ do not live necessarily on the $k–$th order jet space, but involve jet 
spaces of any order. This is equivalent to say that a conceptually complete theory of PDEs is possible only on 
{\it infinite order} jet spaces. A logical consequence of this fact is that objects of the {\it category of 
partial differential equations} are {\it diffieties}, which duly formalize the vague idea of the ``space of all 
solutions'' of a PDE. Diffieties are a kind of infinite dimensional manifolds, and the specific differential 
calculus on them, called {\it secondary calculus}, is a native language to deal with PDEs and especially with 
NPDEs.
\end{quotation}

{Roughly speaking, for each differential equation $\mathcal E$, the basic geometric object 
$\mathcal E^{\infty}$, called ``diffiety'', is introduced. This is the infinitely prolonged equation of the PDE 
$\mathcal E$ or, the space of (pointwise!) formal solutions of $\mathcal E$, and it is endowed with the Cartan 
distribution $\mathcal C$, which is integrable. The dimension of this distribution is equal to the number of 
independent variables of the given PDE, while local solutions of the system correspond to ``integral 
submanifolds'' of $\mathcal C$. This structure is induced by infinite prolongations of PDE, and it is based on 
them. The vector fields on $\mathcal E^{\infty}$ which preserve  $\mathcal C$ are called $\mathcal C-$fields,  
and the infinitesimal symmetries of $\mathcal E$ are defined to be the equivalence classes of 
$\mathcal C-$fields modulo vector fields that are tangent to $\mathcal C.$}

From an algebraic point of view, diffieties are spectra of scalar differential invariants (see examples in 
\cite{Vi}). One can think about them as about infinite-dimensional varieties enabled with a finite dimensional 
involutive Cartan distribution. By means of this distribution, the de Rham algebra on a diffiety can be used to 
define  a spectral sequence (so-called {\it Vinogradov's $\mathcal C-$ spectral sequence}) that encodes 
information on the equation at hand, see \cite{A,KVi,Vi}. The $\mathcal C-$ spectral sequence corresponding to 
the ``empty'' equation, (that is, $\mathcal E^{\infty} = \mathcal J^{\infty}$) is the 
so-called variational bicomplex associated to the bundle $E$, see \cite{A}. 

Let us give a more precise and computable definition of a diffiety. Let $\mathcal E^{\infty}$ be the infinite 
prolongation of a differential equation $\mathcal E$ ({\em i.e.}, $\mathcal E$ is a submanifold of the $k-$th 
jet fibration $\mathcal J^k(\pi)$ of a fibration $\pi : E\mapsto M$).

A diffiety $\mathcal O = (\mathcal O ,C^{\infty}(\mathcal O),\mathcal C(\mathcal O ))$ is a triple consisting 
of a manifold, a function field on it, and a finite-dimensional distribution, that is locally of the form 
$(\mathcal E^{\infty} , C^{\infty} (\mathcal E^{\infty} ),\mathcal C(\mathcal E))$, where 
$\mathcal C(\mathcal E))$ is the restriction of the Cartan distribution on $J^\infty E$ to 
$\mathcal E^{\infty}$. The dimension of $\mathcal C(\mathcal E)$ is called the dimension of 
$\mathcal O$. A smooth map $F:  \mathcal O_1 \mapsto \mathcal O_2$ is called a morphism of diffieties if 
$$dF_x({\mathcal C}_x({\mathcal O}_1))\subset{\mathcal C}_{F(x)}({\mathcal O}_2), \quad x\in {\mathcal O}_1.$$
We note at this point that the notion of a diffiety appears to be very useful also in conceptual definitions of 
non-local symmetries for PDEs, see \cite{KV0,KVi}.

In spite of its ``universality'', ``naturality'' and ``conceptuality'' (the three most popular adjectives used 
by A. M. Vinogradov) these important and useful (and, usually) infinite-dimensional (in a standard sense) 
objects,  may look  ``rather esoteric'' at a first glance 
(to quote Toru Tsujishita in his MR reference to the paper \cite{Vin1984}, but in the op. cit. he immediately 
stresses that, after the paper \cite{Vin1984}, this notion ``now becomes open to everybody.''). We believe
that esoteric they are not: there are some not sufficiently known curious links between the notion of a diffiety 
and other not-so-well-known concepts like Souriau diffeology and Fr\"olicher space structure. In fact, we hope 
to show in this paper that {\em Fr\"olicher spaces} are a very natural arena for the development of the theory 
of diffieties. We can rephrase the definition given above fully rigorously, and we can rather straightforwardly
set up the Vinogradov sequence. 

Fr\"olicher spaces were first described by A. Fr\"olicher in a serie of works in the 1980's (see e.g. 
\cite{F1982,F1983}) and most results of this period are gathered in his book with A. Kriegl \cite{FK}. The  
terminology ``Fr\"olicher space'' can be found for the first time, to our knowledge, in papers by Cherenack (see 
e.g. \cite{Che1998}), and Kriegl and Michor confirmed that these spaces were due to A. Fr\"olicher in \cite{KM}. 
This setting is designed to obtain a safe differential calculus and a safe differential geometry, and the 
presence of charts is not assumed. Smoothness depends directly on tests on a space of functions and on a space 
of paths, that are both assumed to be spaces of smooth maps, under mild conditions for coherence. Refined in 
\cite{Ma2006-3,Wa} by means of the notion of {\em diffeology} due to J.-M. Souriau (see e.g. \cite{MR2016}), 
the technical possibilities in this setting have increased drastically these last few years. Considerations on 
Fr\"olicher spaces can be found in various contexts, see e.g. 
\cite{BN2005,BT2009,B2015,BT2017,C2015,Can2020,DN2007,Ma2006-3,Ma2019,MR2016,Ma2020-3,Nt2005}.  

In this work, written for a proceedings volume in memory of the scientific work of Alexandre Vinogradov, we 
intend to propose a formal description of the interplay of Fr\"olicher spaces and diffieties, as stated in the
penultimate paragraph. We are aware that our formulations 
may not fully fill all the refinements on diffieties that have been produced till now (for example the theory of 
non-local symmetries and coverings, see \cite{KV0}), but this note intends to be a primary base for adaptations 
and discussions. We organize our paper as follows: Section 2 is an introduction to the theory of Fr\"olicher
 spaces and some basic geometric structures on them (tangent vectors, differential forms, etc.); Section 3 is
our proposal for a definition of a diffiety {\em via} Fr\"olicher spaces; Section 4 is both an application of 
the theory and a prelude to further work (we provide some details in the following paragraph);
section 5 contains a final remark.   

On Section 4:  we have explored the initial value problem of the Kadomtsev-Petviashvili (KP) hierarchy and
related non-linear KP equations in several papers, see \cite{ER2013,ERMR,Ma2013,MR2016, MRR}. In 
particular, in \cite{MRR} we have presented several curious equations that correspond to ``KP flows'' posed
on particular algebras equipped with derivations, and in \cite{MRR-jets} we have applied our constructions
to algebras arising from the {\em elementary diffiety} 
$(\mathcal E^{\infty} , C^{\infty} (\mathcal E^{\infty} ),\mathcal C(\mathcal E))$. Can we apply them to
general diffieties as defined in Section 3? In Section 4 we show that the answer is affirmative, provided that 
the diffiety is equipped with a derivation. 


\section{Preliminaries on Fr\"olicher spaces}
\subsection{Fr\"olicher spaces and their diffeology}
\begin{definition} $\bullet$ A \textbf{Fr\"olicher} space is a triple
		$(X,\F,\p)$ such that
		
		- $\p$ is a set of paths $\R\rightarrow X$,
		
		- A function $f:X\rightarrow\R$ is in $\F$ if and only if for any
		$c\in\p$, $f\circ c\in C^{\infty}(\R,\R)$;
		
		- A path $c:\R\rightarrow X$ is in $\p$ (i.e. is a \textbf{contour})
		if and only if for any $f\in\F$, $f\circ c\in C^{\infty}(\R,\R)$.
		
		\vskip 5pt $\bullet$ Let $(X,\F,\p)$ et $(X',\F',\p')$ be two
		Fr\"olicher spaces, a map $f:X\rightarrow X'$ is \textbf{differentiable}
		(=smooth) if and only if one of the following equivalent conditions is fulfilled:
		\begin{itemize}
			\item $\F'\circ f\circ\p\subset C^{\infty}(\R,\R)$
			\item $f \circ \p \subset \p'$
			\item $\F'\circ f \subset  \F$ 
		\end{itemize}
	\end{definition}
	
	Any family of maps $\F_{g}$ from $X$ to $\R$ generates a Fr\"olicher
	structure $(X,\F,\p)$, setting \cite{KM}:
	
	- $\p=\{c:\R\rightarrow X\hbox{ such that }\F_{g}\circ c\subset C^{\infty}(\R,\R)\}$
	
	- $\F=\{f:X\rightarrow\R\hbox{ such that }f\circ\p\subset C^{\infty}(\R,\R)\}.$
	
	We easily see that $\F_{g}\subset\F$. This notion will be useful
	in the sequel to describe in a simple way a Fr\"olicher structure.
	A Fr\"olicher space carries a natural topology, the pull-back topology of $\R$ via $\F$. In the case of
	a finite dimensional differentiable manifold, the underlying topology
	of the Fr\"olicher structure is the same as the manifold topology. In
	the infinite dimensional case, these two topologies differ, in general.
	We have to mention here that the space of smooth functions $\F$ is a vector space, which has a natural 
	Fr\"olicher structure called functional Fr\"olicher structure.
	
	Associated to the notion of Fr\"olicher structure, we find the notion of \textbf{(reflexive) diffeology}. 
	
	\begin{definition} \label{d:diffeology} Let $X$ be a set.
		
		\noindent $\bullet$ A \textbf{p-parametrization} of dimension $p$ 
		on $X$ is a map from an open subset $O$ of $\R^{p}$ to $X$.
		
		\noindent $\bullet$ A \textbf{diffeology} on $X$ is a set $\mathcal{D}$
		of parametrizations on $X$ such that:
		
		\begin{itemize}
\item For each $p\in\N$, any constant map $\R^{p}\rightarrow X$ is in $\mathcal{D}$;
\item \label{d:local} For each arbitrary set of indexes $I$ and family $\{f_{i}:O_{i}\rightarrow X\}_{i\in I}$
			of compatible maps that extend to a map $f:\bigcup_{i\in I}O_{i}\rightarrow X$,
			if $\{f_{i}:O_{i}\rightarrow X\}_{i\in I}\subset\mathcal{D}$, then $f\in\p$.
\item \label{d:compose} For each $f\in\mathcal{D}$, $f : O\subset\R^{p} \rightarrow X$, and 
$g : O' \subset \R^{q} \rightarrow O$, 
			in which $g$ is  
		a smooth map (in the usual sense) from an open set $O' \subset \R^{q}$ to $O$, we have $f\circ g\in\p$.
		\end{itemize} 
		
		\vskip 6pt If $\mathcal{D}$ is a diffeology on $X$, then $(X,\mathcal{D})$ is
		called a \textbf{diffeological space} and, if $(X,\mathcal{D})$ and $(X',\mathcal{D}')$ are two 
		diffeological spaces, 
		a map $f:X\rightarrow X'$ is \textbf{smooth} if and only if $f\circ\mathcal{D}\subset\mathcal{D}'$. 
	\end{definition} 
	
	The notion of a diffeological space is due to J.M. Souriau, see \cite{Sou}; a previous but not fully 
	equivalent notion is due to Chen, see \cite{Chen} and \cite{St} for a very careful comparison.
	A comprehensive exposition of basic concepts can be found in   \cite{Igdiff}.  The category of diffeological 
	spaces is very large, and it contains many different pathological examples even if it enables a very 
	easy-to-use framework for infinite dimensional objects. Therefore, the category of Fr\"olicher spaces as a 
	subcategory
	of this category (see \cite{Ma2006-3,Wa,St,Ma2020-3}) with less technical problems may be useful.  The first 
	steps of the comparison were published in \cite{Ma2006-3}; the reader can 
	also see \cite{Ma2013,MR2016,Wa} for extended expositions.
	In particular, it is explained in \cite{MR2016} that 
	{\em Diffeological, Fr\"olicher and Gateaux smoothness are the same notion if we 
		restrict ourselves to a Fr\'echet context,} in a sense that we explain here. For this, we first need to 
		analyze how we generate a Fr\"olicher or a diffeological space, that is, how we implement a Fr\"olicher 
		or a diffeological structure on a given set $X.$ 
		
	If $(X,\F, \C)$ is a Fr\"olicher space, we 
	define a natural diffeology on $X$ by using the following family
	of maps $f$ defined on open domains $D(f)$ of Euclidean spaces, see \cite{Ma2006-3}:
	$$
	\mathcal{D}_\infty(\F)=
	\coprod_{p\in\N^*}\{\, f: D(f) \rightarrow X\, | \, D(f) \hbox{ is open in } \R^p 
	\hbox{ and } \F \circ f \in C^\infty(D(f),\R) \}\; .$$
	Now we can easily show the following:
	
	\begin{proposition} \label{fd} \cite{Ma2006-3}
		Let$(X,\F,\p)$
		and $(X',\F',\p')$ be two Fr\"olicher spaces. A map $f:X\rightarrow X'$
		is smooth in the sense of Fr\"olicher if and only if it is smooth for
		the underlying diffeologies $\mathcal{D}_\infty(\F)$ and $\mathcal{D}_\infty(\F').$
	\end{proposition}
	
	Thus, Proposition \ref{fd} and the foregoing remarks imply that 
	the following implications hold:
	\vskip 12pt
	
	\begin{tabular}{ccccc}
		smooth manifold  & $\Rightarrow$  & Fr\"olicher space  & $\Rightarrow$  & diffeological space
	\end{tabular}
	{
		\vskip 12pt 
		The reader is referred to the Ph.D. thesis \cite{Wa} for a deeper analysis of these implications.
	A more complete and up-to-date exposition, based on infinite dimensional examples, is actually under 
	preparation in \cite{GMW}, but let us make some short precisions for the reader. 
	We follow
	\cite[Section 3]{F1982} closely: as a consequence of Boman's theorem, see \cite{KM}, the triple 
	$(\R^n , C^{\infty}(\R^n,\R) , C^{\infty}(\R,\R^n))$ is a 	Fr\"olicher structure on $\R^n$. More generally,
	if $V$ is a finite-dimensional paracompact smooth manifold, $(V,C^{\infty}(V,\R) , C^{\infty}(\R,V))$ is a 
	Fr\"olicher structure on $V$ and the smooth (in the standard sense) maps between two such manifolds are
	precisely the smooth maps in the Fr\"olicher sense. We can replace $\R^n$ for a Fr\'echet space $E$ in our
	first example, and we can replace $V$ for a Fr\'echet manifold $V'$ in our second example, {\em provided
		that} $V'$ is a paracompact topological space and it is modelled on a Fr\'echet space $E'$ that 
	satisfies the following property: for each neighbourhood $U$ of $0 \in E'$ there exists a smooth function 
	$f :U \rightarrow \R$ such that $f(0) = 1$ and $f(x) =0$ for $x \not \in U$.
	
	\subsection{Tangent spaces, differential forms, diffeomorphisms and all that}
	
	On Fr\"olicher spaces, we can define internal tangent cones ${}^iT_xX,$ see e.g. \cite{Ma2013}. We mention 
	that, diffeologically, internal tangent spaces are defined in a different way, see e.g. \cite{CW}, but the 
	approach that we present here is 
	\begin{itemize}
		\item coherent with the definition of the kinematic tangent space of a manifold in the $c^\infty$ 
		category studied in \cite{KM}
		\item and chosen as an interesting definition for tangent space in an applied setting 
		\cite{GW2021,Ma2013,MR2016}.
	\end{itemize}
		For each $x\in X,$ we consider $$C_{x}=\{c \in C^\infty(\R,X)| c(0) = x\}$$ 
	and take the equivalence relation $\mathcal{R}$ given by $$c\mathcal{R}c' \Leftrightarrow \forall 
	f \in C^\infty(X,\R), \quad \partial_t(f \circ c)|_{t = 0} = \partial_t(f \circ c')|_{t = 0}\; .$$
	The internal tangent cone at $x$ is the quotient
	$$^iT_xX = C_x / \mathcal{R}.$$ If $X = \partial_tc(t)|_{t=0} \in {}^iT_X, $ we define the simplified 
	notation  $$Df(X) = \partial_t(f \circ c)|_{t = 0}\; .$$
	Under these constructions, we can define the total space of internal tangent cones 
	$$ {}^{i}TX = \coprod_{x \in X} {}^{i}T_xX$$ 
	with canonical projection $\pi: u \in {}^{i}T_xX \mapsto x.$ A Fr\"olicher structure can be defined on it. 
	We note that in general, the fibration $\pi$ on ${}^{i}TX$ may have non-isomorphic fibers that are 
	all cones but not vector spaces (an example appears in \cite{Ma2020-3}). 
	
	Differential forms on $X$ and a de Rham differential also exist in this setting. It is easier to 
	understand their construction under the diffeology viewpoint:
	
	\begin{definition} \cite{Sou}
		Let $(X,\mathcal{D})$ be a diffeological space and let $V$ be a vector space equipped with a 
		differentiable structure. 
		A $V-$valued $n-$differential form $\alpha$ on $X$ (noted $\alpha \in \Omega^n(X,V))$ is a map 
		$$ \alpha : \{p:O_p\rightarrow X\} \in \mathcal{D} \mapsto \alpha_p \in \Omega^n(O_p;V)$$
		such that 
		
		$\bullet$ Let $x\in X.$ $\forall p,p'\in \mathcal{D}$ such that $x\in Im(p)\cap Im(p')$, 
		the forms $\alpha_p$ and $\alpha_{p'}$ are of the same order $n.$ 
		
		$\bullet$ Moreover, let $y\in O_p$ and $y'\in O_{p'}.$ If $(X_1,...,X_n)$ are $n$ germs of paths in 
		$Im(p)\cap Im(p'),$ if there exists two systems of $n-$vectors $(Y_1,...,Y_n)\in (T_yO_p)^n$ and 
		$(Y'_1,...,Y'_n)\in (T_{y'}O_{p'})^n,$ if $p_*(Y_1,...,Y_n)=p'_*(Y'_1,...,Y'_n)=(X_1,...,X_n),$
		$$ \alpha_p(Y_1,...,Y_n) = \alpha_{p'}(Y'_1,...,Y'_n).$$
		
	We note by $$\Omega(X;V)=\oplus_{n\in \mathbb{N}} \Omega^n(X,V)$$ the set of $V-$valued differential forms.  
	\end{definition}
	
	We feel we need to make two remarks:
	
	$\bullet$ If do not exist $n$ linearly independent vectors $(Y_1,...,Y_n)$
	defined as in the last point of the definition, $\alpha_p = 0$ at $y.$
	
	$\bullet$ Let $(\alpha, p, p') \in \Omega(X,V)\times \mathcal{D}^2.$ 
	If there exists $g \in C^\infty(D(p); D(p'))$ (in the usual sense) 
	such that $p' \circ g = p,$ then $\alpha_p = g^*\alpha_{p'}.$

	\begin{proposition}
		The set $\p(\Omega^n(X,V))$ made of maps $q:x \mapsto \alpha(x)$ from an open subset $O_q$ of a 
		finite dimensional vector space to $\Omega^n(X,V)$ such that for each $p \in \p,$ $$\{ x \mapsto 
		\alpha_p(x) \} \in C^\infty(O_q, \Omega^n(O_p,V)),$$
		is a diffeology on $\Omega^n(X,V).$  
	\end{proposition}
	
	Working with plots of the diffeology, we can define the wedge product and the exterior  differential 
	of differential forms, which are (diffeologically) smooth and have the same properties as in the standard
	case.
	
	We need to introduce one last construction. Again following \cite{Ma2020-3}, we let $Diff(X)$ be the group 
	of diffeomorphisms of the Fr\"olicher space $X$. This is a well-defined object, it can be equipped
	with a diffeology, and its internal tangent space at the identity is a diffeological vector space 
	that defines (see \cite[Section 2]{Ma2020-3}) a restricted tangent space on $X$ called diff-tangent space 
	and noted by $ {}^{d}TX$. The space $ {}^{d}TX$ is a Fr\"olicher subspace of $ {}^{i}TX$ and it is also a 
	Fr\"olicher vector bundle. Hereafter we assume that $Diff(X)$ is a \textbf{Fr\"olicher Lie group}, that is, 
	the tangent space at the identity has a natural diffeology for which the Lie bracket is smooth. This 
	condition is not automatically fulfilled, and the study \cite{Les}, completed in the paper \cite{Lau2011}, 
	shows that	this happens only on a ``restricted" class of examples. However it is only through this setting 
	that one can, to our knowledge, define a Lie bracket on vector fields that are identified as elements of the 
	Lie algebra of $Diff(X).$

\section{Diffieties and their Fr\"olicher structure}
	
	We recall from \cite[p. 7, 191]{Vi} and Section 1, that a {\em diffiety} is, classically, a (finite or 
	infinite dimensional) manifold ${\mathcal O}$ that is locally of the form $\mathcal{E}^{\infty}$, where 
	$\mathcal{E}^{\infty}$ denotes an infinitely prolonged equation that fibers over an $n$-dimensional manifold 
	of ``independent variables''. We refer the reader to \cite{KVi} for information on equation manifolds. 
	This definition means, in particular, that ${\mathcal O}$ must be equipped with an $n$-dimensional 
	involutive distribution that coincides (locally) with the Cartan distribution of $\mathcal{E}^{\infty}$. 
	
	We now propose a very general definition of a diffiety that does not use topology explicitly. Let 
	$\mathcal{O}$ be a 	set, and let  $(\mathcal{O}, \F({\mathcal O}), \p(\mathcal{O}))$ be a Fr\"olicher 
	structure. Since $\p(\mathcal{O})$ is generated by $\F(\mathcal{O}),$ we remark that it is redundant to 
	mention it.
	
	\begin{definition} 
		We consider a Fr\"olicher space $(\mathcal{O}, \F({\mathcal O}), \p(\mathcal{O}))$. 
		A \textbf{Cartan distribution} $\C(\mathcal{O})$ on $\mathcal{O}$ is a finite-dimensional vector 
		subbundle of ${}^dTX$ that is involutive, that is, there is a Lie subalgebra $\mathcal{H}$ of the Lie 
		algebra of $Diff(\mathcal{O})$ such that for all $p \in \mathcal{O}$, 
		$$\C(\mathcal{O})_p = \{ X(p): X \in \mathcal{H} \}\; .$$  
		A \textbf{diffiety} is a triple $(\mathcal{O}, \F({\mathcal O}), \p(\mathcal{O}))$.
	\end{definition}
	
	This definition includes the particular case of $\mathcal{O} =\mathcal{E}^{\infty}$ if all prolongations of 
	$\mathcal{E}$ satisfy natural conditions as in the standard case of \cite{A,KVi}. Indeed, let us assume that 
	$\mathcal{E}^{\infty}$ 
	is contained in the infinite jet bundle $J^\infty(E)$ for some fiber bundle $E \rightarrow M$. The bundle
	$J^\infty(E)$ is the inverse limit of the sequence of finite-dimensional jet bundles $J^k(E)$, $k\geq 0$,
	and we have observed that these manifolds have natural Fr\"olicher structures. Now, the category of 
	Fr\"olicher spaces and smooth maps is closed under inverse limits (see e.g., \cite{FK}), and therefore we 
	obtain	that 
	$J^\infty(E)$ is a Fr\"olicher space. Since $\mathcal{E}^{\infty} \subseteq J^\infty(E)$, we conclude that 
	$\mathcal{E}^{\infty}$ has a 
	natural Fr\"olicher structure. It follows from the standard geometry reviewed for example in \cite{A,KVi}, 
	that functions over $\mathcal{E}^{\infty}$ that are Fr\"olicher-smooth are precisely the standard smooth 
	functions on $\mathcal{E}^{\infty}$. Thus, the classical example of a diffiety (what Vinogradov calls an 
	{\em elementary diffiety} in \cite[P. 191]{Vi}) {\em is} a diffiety in our sense.
	
	\smallskip
	
We now show that within this framework, we can build the \textbf{Vinogradov sequence} along the lines of 
\cite{Vi1984,Vi}:
the algebra of differential forms
$$ \Omega^*(\mathcal{O}) = \sum_{k \in \N} \Omega^*(\mathcal{O})$$ is a graded differential algebra for the de 
Rham differential operator. Let us define 
$$ C\Omega^k(\mathcal{O}) = \left\{\alpha \in \Omega^k \, | \, \alpha|_{\mathcal{C}(\mathcal{O}) }= 0\right\},$$
$$ \mathcal{C}\Omega(\mathcal{O}) = \sum_{k \in \N}  \mathcal{C}\Omega^k(\mathcal{O})$$
	and $$\mathcal{C}^l \Omega(\mathcal{O}) = \mathcal{C}\Omega(\mathcal{O})^{\wedge l}.$$
	Since $\mathcal{C}(\mathcal{O})$ is involutive, $\mathcal{C}(\mathcal{O})$ is a differential ideal, and so 
	is $\mathcal{C}^l(\mathcal{O})$ for $l \in \N^*.$ 
Therefore we can define 
$$E^{p,q}_0
 = \frac{\mathcal{C}^p\Omega^{p+q}(\mathcal{O})}{\mathcal{C}^{p+1}\Omega^{p+q}(\mathcal{O})},$$
 $$ E^{p,q}_{r+1} = H^*( E^{p,q}_{r})$$
 and $$\mathcal{C} E(\mathcal{O}) = \left\{ E_r^{p,q}\right\}$$ equipped with the restriction/extension of the 
 de Rham differential to the corresponding spaces. This is the spectral 
 sequence we sought.

 	\section{KP hierarchy on a diffiety}
In this section we consider the KP hierarchy. As explained in Section 1, we present this section as an advance
of \cite{MRR-jets}, where we study differential equations posed on {\em elementary} diffieties. Here, for the
sake of brevity we simply assume that there is a non-trivial derivation $D$ on $\F(\mathcal{O})$ that is 
smooth (in \cite{MRR-jets} we use explicit derivations). We denote by  $A$ any Fr\"olicher subalgebra of 
$\mathcal{F}(\mathcal{O)}$ such that $D\, A \subset A.$

Let $\xi$ be a formal variable not in $A$. The {\em algebra of symbols} over $A$ is the vector space
\[
\Psi_{\xi}(A) = \left \{ P_{\xi} = \sum_{\nu \in {\bf Z}} a_{\nu} \, \xi^{\nu} : a_{\nu}
\in A \, , \;  a_{\nu} = 0 \mbox{ for } \nu \gg 0 \right \} \;
\]
equipped with the associative multiplication $\circ\,$, where
\begin{equation}  \label{alfa}
	P_{\xi} \circ Q_{\xi} = \sum_{k \geq 0} \frac{1}{k !} \, \frac{\partial^{k}P_{\xi}}
	{\partial \xi^{k}} \, D^{k} Q_{\xi} \; ,
\end{equation}
with the prescription that the multiplication on the right hand
side is the standard multiplication of Laurent series in $\xi$
with coefficients in $A$. The algebra $A$ is included in $\Psi_\xi(A)$. 

The {\em algebra of formal pseudo-differential operators} over $A$ is the vector space
\[
\Psi (A) = \left \{ P = \sum_{\nu \in \mathbb{Z}} a_{\nu} \,
D^{\nu} : a_{\nu} \in A \, , \; a_{\nu} = 0 \mbox{ for } \nu \gg 0
\right \} \; ,
\]
in which a multiplication $\circ$ is defined so that the map
\[
\sum_{\nu \in \mathbb{Z}} a_{\nu} \, \xi^{\nu} \; \mapsto \;
\sum_{\nu \in \mathbb{Z}} a_{\nu} \, D^{\nu}
\]
from $\Psi_{\xi}(A)$ to $\Psi (A)$ is an algebra homomorphism. The
algebra $\Psi (A)$ is associative; it becomes a Lie algebra over $K$ if we define, as usual,
\begin{equation}
	[ P , Q ] = P \circ Q - Q \circ P \; .  \label{liebrack}
\end{equation}

The {\em order} of $P \neq 0 \in \Psi (A)$ is $N$ if $a_{N} \neq
0$ and $a_{\nu} = 0$ for all $\nu > N$, and order($P$) = $-
\infty$ if $P = 0$. If $P$ is of order $N$, the coefficient
$a_{N}$ is called the {\em leading term} of $P$.

We collect some properties of $\Psi (A)$ for completeness.

\begin{lemma} \label{lieprop}  
	\begin{enumerate}
		\item If $P$ and $Q$ are formal pseudo--differential operators over $A$, and the leading terms
		of $P$, $Q$ are not divisors of zero in $A$, then ${\rm order}(P \circ Q) = {\rm order}(P)
		+ {\rm order}(Q)$, and
		${\rm order}([P,Q]) \leq {\rm order}(P) + {\rm order}(Q) -1$. In particular, if $A$ is an integral 
		domain,
		the ring $\Psi(A)$ does not have zero divisors.
		\item Every non--zero formal pseudo--differential operator $P$ for which its leading term
		is invertible has an inverse in $\Psi (A)$.
	\end{enumerate}
\end{lemma}

The following observation is at the basis of all our analysis: the Lie algebra $\Psi(A)$ admits a left 
$A$--module direct sum decomposition
\begin{equation} \label{desc}
	\Psi(A) = {\mathcal I}_{A} \oplus {\mathcal D}_{A}   \; ,
\end{equation}
in which ${\mathcal D}_{A}$ is the Lie subalgebra of all {\em
	differential} operators of order greater or equal to zero, and
${\mathcal I}_{A}$ is the Lie subalgebra of {\em integral} operators,
that is, the set of all formal pseudo-differential operators in
$\Psi(A)$ of order at most $-1$. It is known since \cite{M1,M2,M3} that there exist formal
Lie groups $G(\Psi(A))$, $G_+({\mathcal D}_{A})$, and $G_{-}({\mathcal
	I}_{A})$ with Lie algebras $\Psi(A)$, $\mathcal{D}_A$, and
$\mathcal{I}_A$ respectively, such that
\begin{equation} \label{desc1}
	G(\Psi(A)) = G_{-}({\mathcal I}_{A}) \cdot G_{+}({\mathcal D}_{A})\,.
\end{equation}
In order to explain what these groups are, we need to specialize the algebra $A$ considered in the 
previous subsection. Let $R$ be a fixed commutative algebra over a field
$K$ equipped with a derivation $D$. We take as $A$ the ring of
formal power series over $R$ in an infinite number of variables
$\tau_1, \tau_2, \cdots.$ Certainly, it is not necessary to make this
particular choice in the classical algebraic theory of KP (see for
instance \cite{D,MJD2000}) but, unless we equip $A$ with a further structure as in Demidov 
\cite{Dem1995,Dem1998} and specially \cite{ERMR}, taking $A$ as a ring of power series
 is crucial for the explicit construction of the groups $G(\Psi(A))$, $G_+({\mathcal D}_{A})$ and
$G_{-}({\mathcal I}_{A})$. 

\begin{remark}
	For the benefit of the reader, we recall the construction of $A$ following Bourbaki \cite[p. 454-457]{B}.
	We consider a countable set of indices $I,$ and we take $T$ as the
	additive monoid of all sequences of natural numbers $t=(n_i)_{i\in
		I}$ such that $n_{i}=0$ except for a finite number of indices. A
	{\em formal power series} is a function $u$ from $T$ to $R$,
	$u=(u_t)_{t\in T}$. Consider an infinite number of formal
	variables $\tau_i$, $i \in I$ and set
	$\tau=(\tau_1,\tau_2,\cdots)$. Then,  the formal power series $u$
	is also written as $u=\sum_{t\in T}u_t\tau^t$ in which $\tau^t =
	\tau_1^{n_1} \tau_2^{n_2}\cdots$. We say that $u_t\in R$ is a
	coefficient and that $u_t\tau^t$ is a term. \\
	Operations on the ring $A$ are defined in an usual manner: If
	$u=(u_t)_{t\in T},$ $v=(v_t)_{t\in T},$ then we set
	$$u+v=(u_t+v_t)_{t\in T} \quad \mbox{ and } \quad uv=w,$$ in which
	$w=(w_t)_{t\in T}$ and
	$$w_t=\sum_{p,q\in T\atop{p+q=t}} u_pv_q.$$
	It is shown in \cite[p. 455]{B} that this multiplication is well defined, and that $A$ equipped with
	these two operations is a commutative algebra with unit. The derivation $D$ on $R$ extends to a derivation
	on $A$ via $$Du=\sum_{t\in T}(Du_t)\tau^t.$$
	We also define the \textit{order} of a power series. Let $u\in A,$
	$u\neq0.$ Let us write $u=\sum_{t\in T}u_t\tau^t$, and if $t =
	(n_i)_{i \in I}$, let us set $\mid t\mid= \sum n_i$. The terms
	$u_t\tau^t$ such that $\mid t\mid=p$ are called {\em terms of
		total degree} $p.$ The formal power series $u_p$ whose terms of
	total degree $p$ are those of $u,$ and whose other terms are zero,
	is called {\em the homogeneous part of $u$ of degree $p.$} The
	series $u_0,$ the homogeneous part of $u$ of degree $0$, is
	identified with an element of $R$ called the constant term of $u.$
	For a formal series $u\neq 0,$ the least integer $p\geq 0$ such
	that $u_p\neq0$ is called the {\em order} of $u,$ and it is
	denoted by $ord_t(u)$, and we extend this definition to the case
	$u=0$ setting $ord_t(0) = \infty$ (see \cite[p. 457]{B}). The
	following properties hold: if $u,v$ are formal power series
	different from zero, then
	$$ord_t(u+v)\geq inf(ord_t(u),ord_t(v))\; , \quad \mbox{ if } u+ v \neq 0 ,$$
	$$ord_t(u+v)=inf(ord_t(u),ord_t(v))\; , \quad \mbox{ if } ord_t(u)\neq ord_t(v) ,$$
	$$ord_t(uv)\geq ord_t(u)+ord_t(v)\; , \quad \mbox{ if } u v \neq 0 .$$
\end{remark}

\vspace{0.8cm}

Following Mulase, \cite{M3}, we now define the space
\begin{equation} \label{aldef}
	\widehat{\Psi}(A) = \left\{ P= \sum_{\alpha \in {\mathbb{Z}}}
	a_{\alpha}\,D^{\alpha} : a_{\alpha} \in A \mbox { and } \exists
	C\in\mathbb{R}^+ , N\in\mathbb{Z}^+ \mbox{ so that }
	ord_t(a_\alpha) > C\,\alpha - N \ \forall\ \alpha \gg 0 \right\}
\end{equation}
and the subspace
\begin{equation}
	\widehat{\mathcal D}_{A} = \left\{ P= \sum_{\alpha \in \mathbb{Z}}
	a_{\alpha}\,D^{\alpha} : P \in\widehat{\Psi}(A) \mbox{ and }
	a_\alpha=0 \mbox { for } \alpha <0 \right\} \; .
\end{equation}

The definition of order explained in the last remark implies that
$A$ and $\Psi(A)$ are contained in $\widehat{\Psi}(A)$. The
operations on $\widehat{\Psi}(A)$ are natural extensions of the
operations on $\Psi(A)$. As mentioned by Mulase in \cite{M3} and proved explicitly in \cite{ER2013, ERMR}, the 
following holds:
\begin{lemma}
	The space $\widehat{\Psi}(A)$ has a natural algebra structure, and
	$\widehat{\mathcal D}_{A}$ is a subalgebra of $\widehat{\Psi}(A)$.
\end{lemma}

\begin{definition}
	Let $\mathcal{K}$ be the ideal of $A$ generated by $t_1, t_2,
	\cdots$. If $P \in \widehat{\Psi}(A)$, we denote by $P\vert_{t=0}$
	the equivalence class $P\, mod\; \mathcal{K}$, and we identify it
	with an element of $\Psi(A)$. We also set $G_A = 1 +
	\mathcal{I}_A$, and we define the spaces
	\[
	\widehat{\Psi}(A)^{\times} = \{ P \in \widehat{\Psi}(A) :
	P\vert_{t=0}\in G_A  \}
	\]
	and
	\[
	\widehat{\mathcal D}_{A}^{\times} = \{ P \in \widehat{\mathcal D}_{A} :
	P\vert_{t=0}=1 \} \; .
	\]
\end{definition}

The most important fact about these two spaces is the following result proven by Mulase in \cite{M3}, and 
reviewed
in full detail in \cite{ER2013} and \cite{ERMR}.

\begin{proposition} \label{gr}
	The spaces $\widehat{\Psi}(A)^{\times}$ and $\widehat{\mathcal D}_{A}^{\times}$ are formal Lie groups: 
	each element $P$ in $\widehat{\Psi}(A)^{\times}$ and $\widehat{\mathcal D}_{A}^{\times}$
	has an inverse of the form
	\[
	P^{-1} = \sum_{n \geq 0} (1 - P)^{n} \; .
	\]
	Moreover, we have:
	$G(\Psi(A)) = \widehat{\Psi}(A)^{\times}$,
	$G_+(\mathcal{D}(A)) = \widehat{\mathcal D}_{A}^{\times}$, and
	$G_-(\mathcal{I}_A) = 1 + {\mathcal I}_{A}$.
\end{proposition}


\begin{theorem} \label{mu2}
	For any $U \in G({\Psi}(A))$ there exist unique $W \in
	G_-(\mathcal{I}_A)$ and $Y \in G_+({\mathcal D}_{A})$ such that
	\[
	U = W^{-1}\,Y \; .
	\]
	In other words, there exists a unique global factorization of the formal Lie group
	$G({\Psi}(A))$ as a product group,
	\[
	G({\Psi}(A)) = G_-(\mathcal{I}_A) \, G_+({\mathcal D}_{A}) \; .
	\]
\end{theorem}

\noindent This theorem is proven in \cite{M3}, see also the later papers \cite{ER2013,ERMR} for full proofs.

\smallskip

The {\em Kadomtsev-Petviashvili (KP) hierarchy} on $\mathcal{O}$ reads
\begin{equation} \label{lolo}
	\frac{d L}{d t_{k}} = \left[ (L^{k})_{+} , L \right]\; , \quad
	\quad k \geq 1 \; ,
\end{equation}
with initial condition $L(0)=L_0  \in \Psi^1 (R)$. The dependent
variable $L$ is chosen to be of the form
$$L = D + \sum_{\alpha \leq -1 } u_\alpha\, D^\alpha \in {\Psi}^1(A_t) \; .$$
Standard references on (\ref{lolo}) are \cite{D,MJD2000,M1,M2,M3}. The
following result 
gives a solution to the Cauchy problem for the KP hierarchy (\ref{lolo}).

\begin{theorem} \label{KPcentral}
	Consider the KP hierarchy
	with initial condition $L(0)=L_0$. 
	\begin{enumerate}
		\item There exists a pair $(S,Y) \in  G_-(\mathcal{I}_A) \times \widehat{\mathcal D}_{A}^{\times}$  such that the unique solution to Equation
		$(\ref{lolo})$ with $L(0)=L_0$ is
		\begin{eqnarray*}
			L(t)=Y\,L_0\,Y^{-1} = S L_0 S^{-1} \; .
		\end{eqnarray*}
		\item The pair $(S,Y)$ is uniquely determined by the decomposition problem $$exp\left(\sum_{k \in \N}
		t_kL_0^k\right) = S^{-1}Y.$$
	\item The solution operator $L$ is smoothly dependent on the variable $t$ and on the initial value $L_0.$
		This means that the map
		$$ (L_0,s) \in (D + \Psi^{-1}(A))\times T \mapsto \sum_{n \in \N} \left( \sum_{|t|=n}[L(s)]_t\right) \in 
		(D + \Psi^{-1}(R))^\N $$
		is smooth, in which $s \in \cup_{n \in \N}\K^n$ and this set is equipped with the structure of locally 
		convex topological space given by the inductive limit.
	\end{enumerate}
\end{theorem}

The existence of an algebraic decomposition as in Parts (1) and
(2) of Theorem \ref{KPcentral} appears already in Mulase's seminal
papers \cite{M1,M3}, and a formal solution to (\ref{lolo}) as in Part
(1) is in \cite{ER2013}. The richness of Theorem \ref{KPcentral}
steams from the fact that we pose the KP equations (\ref{lolo}) in
the Fr\"olicher algebra $\Psi(A)$  using
an analytically rigorous factorization in Fr\"olicher  groups
of the infinite-dimensional Fr\"olicher 
group $G({\Psi}(A))$. For all details, we refer to \cite{ERMR,MR2016}.

	\section{Outlook}
As anticipated,	we have a more concrete example of Fr\"olicher structure on a classical elementary diffiety in 
the work in progress \cite{MRR-jets}; in \cite{MRR-jets} we also investigate analogues to the KP hierarchy and 
equations very explicitly. The definition of a diffiety given here, plus Section 5, compels us to study 
non-linear equations ``of KP type'' in full detail in this global setting. We hope to do so in the near future. 

\bibliographystyle{amsalpha}

\begin{thebibliography}{A}

\bibitem{A} Anderson, I.M.; Introduction to the variational 
		bicomplex. \emph{Contemporary Mathematics} \textbf{132}  51--73 (1992).
\bibitem{BN2005} Batubenge, A.; Ntumba, P.; On the way to Fr\"olicher Lie groups
		\textit{Quaestionnes mathematicae}  \textbf{28} no1, 73--93 (2005)
\bibitem{BT2009} Batubenge, A.; Tshilombo, H.
		Topologies and smooth maps on initial and final objects in the category of Fr\"olicher spaces. 
		\emph{Demonstr. Math.} {\bf 42}, No. 3, 641-655 (2009). 
\bibitem{B2015} Batubenge, A.; 	A survey on Fr\"olicher spaces. \emph{Quaest. Math.}
		{\bf 38}, no. 6, 869-884 (2015).
\bibitem{BT2017} Batubenge, A.; Tshilombo, H.;
		Finsler metric topology coincides with Fr\"olicher topology. 
		\emph{Balkan J. Geom. Appl.} {\bf 22}, no. 2, 1-12 (2017)
\bibitem{B} Bourbaki, N.; {\em Algebra I. Chapters 1--3.} Elements of Mathematics (Berlin).
		Springer-Verlag, Berlin, 1998.
\bibitem{C2015} Canarutto, D.; Fr\"olicher-smooth geometries, quantum jet bundles and BRST symmetry.
		\emph{J. Geom. Phys.} \textbf{88}, 113--128 (2015). 
\bibitem{Can2020} Canarutto, D.; \emph{Gauge Field Theory in Natural Geometric Language: A revisitation 
		of mathematical notions of quantum physics} Oxford University Press (2020)
\bibitem{Chen}  Chen, K.T.; \emph{Iterated Path Integrals}. Bulletin AMS \textbf{83} , 831--879 (1977).
\bibitem{Che1998} Cherenack, P.; Applications of Fr\"olicher spaces to cosmology. 
              \emph{	Ann. Univ. Sci. Budap. Rolando E\"otv\"os, Sect. Math.} \textbf{41}, 63-91 (1998). 
\bibitem{CW}  Christensen, J.D.; Wu, E.; Tangent spaces and tangent bundles for diffeological spaces;  
		{\it Cahiers de Topologie et G\'eom\'etrie Diff\'erentielle},  \textbf{LVII}, 3-50 (2016)
\bibitem{Dem1995} Demidov, E. E., On the Kadomtsev-Petviashvili hierarchy with a noncommutative 
		             timespace, Funct. Anal. Appl., 29, 2, 131-133, (1995)
\bibitem{Dem1998} Demidov, E. E., Noncommutative deformation of the Kadomtsev-Petviashvili hierarchy. 
	             In ``Algebra. 5, Vseross. Inst. Nauchn. i Tekhn. Inform. (VINITI)'', Moscow, 1995. (Russian), 
	            J. Math. Sci. (New York), 88, 4, 520-536, (1998)
\bibitem{DN2007} Dugmore, B.; Ntumba, P. P. On tangent cones of Fr\"olicher spaces. 
		\emph{Quaest. Math.} \textbf{30} no. 1, 67-83 (2007). 
\bibitem{D} Dickey, L.A.; \textit{Soliton equations and Hamiltonian systems, second edition} (2003).
\bibitem{ER2013} Eslami Rad, A.; Reyes, E. G.; The Kadomtsev-Petviashvili hierarchy and the Mulase factorization 
        of formal Lie groups {\it J. Geom. Mech.} 5 no 3 (2013), 345--363.
\bibitem{ERMR} Eslami Rad, A.; Magnot, J.-P.; Reyes, E. G.; The Cauchy problem of the Kadomtsev-
		Petviashvili 
		hierarchy with arbitrary coefficient algebra. {\it J. Nonlinear Math. Phys.} 24:sup1 (2017), 103--120.
\bibitem{F1982}	Fr\"olicher, A.; Smooth structures. in:
		\emph{Category theory. Applications to algebra, logic and topology, Proc. int. Conf., Gummersbach 1981, 
		Lect. Notes Math.} {\bf 962}, 69-81 (1982). 
\bibitem{F1983} Fr\"olicher, A.; Cartesian closed categories and analysis of smooth maps in:  
		\emph{Categories in continuum physics, Lect. Notes in Math.} \textbf{1174}J 43--51 (1983)
\bibitem{FK} Fr\"olicher, A; Kriegl, A; {\it Linear spaces and differentiation theory} Wiley series in Pure and 
        Applied Mathematics, Wiley Interscience (1988) 
\bibitem{GW2021}
        Goldammer, N. Welker, K.; Optimization in diffeological spaces. 
        \emph{Proc. Appl. Math. Mech.} {\bf 21} no S1 e202100260 (2021)
\bibitem{GMW} Goldammer, N.; Magnot, J-P.; Welker, K.; On diffeologies in infinite dimensional geometry and 
        shape analysis (preliminary title). In preparation.
\bibitem{Igdiff} Iglesias-Zemmour, P.; \textit{Diffeology} Mathematical Surveys and Monographs
			\textbf{185} (2013).
\bibitem{KV0} Krasil'shchik, I.S.; Vinogradov, A.M.; Nonlocal
			trends in the geometry of differential equations: Symmetries,
			conservation laws and B\"{a}cklund transformations. {\em Acta Appl. Math.} 15 (1989), 161--209.
\bibitem{KVi} I.S. Krasil'shchik, and A.M. Vinogradov (Eds.) \emph{Symmetries
			and conservation laws for differential equations of mathematical
			physics}  Translations of Mathematical Monographs Vol. \textbf{182}, AMS, Providence(1999).
\bibitem{KM} Kriegl, A.; Michor, P.W.; \textit{The convenient setting
			for global analysis} Math. surveys and monographs \textbf{53}, American
		Mathematical society, Providence, USA. (2000)
\bibitem{Lau2011} Laubinger, M.; A Lie algebra for Fr\"olicher groups \textit{Indag. Math.} 
			\textbf{21} no 3-4, 156--174 (2011) 
\bibitem{Les} Leslie, J.; On a Diffeological Group Realization of certain Generalized symmetrizable Kac-
		Moody Lie Algebras \textit{J. Lie Theory} \textbf{13} (2003), 427-442.
\bibitem{Ma2006-3} Magnot, J-P.; Diff\'eologie sur le fibr\'e d'holonomie d'une connexion en dimension 
		infinie
		\textit{C. R. Math. Acad. Sci., Soc. R. Can.} \textbf{28}, no. 4, 121-128 (2006)
\bibitem{Ma2013} Magnot, J-P.; Ambrose-Singer theorem on diffeological bundles and complete 
		integrability of KP equations. {\em Int. J. Geom. Meth. Mod. Phys.} {\bf 10}, no 9 Article ID
		1350043 (2013)
\bibitem{Ma2019} Magnot, J-P.; Remarks on the geometry and the topology of the loop spaces $H^s(S^1,N),$ for $s 
        \leq 1/2.$ \emph{Int. J. Maps Math.} {\bf 2}  no. 1, 14–37 (2019).
\bibitem{MR2016} Magnot, J-P.; Reyes, E. G.; Well-posedness of the Kadomtsev-Petviashvili hierarchy, 
		Mulase factorization, and Fr\"olicher Lie groups.  \textit{Ann. H. Poincar\'e} { \bf 21}, No. 6, 
		1893--1945 (2020).
\bibitem{Ma2020-3} Magnot, J-P.; 
		On the differential geometry of numerical schemes and weak solutions of functional equations. 
		\emph{Nonlinearity} {\bf 33}, No. 12, 6835-6867 (2020)
\bibitem{MRR} Magnot,J.-P.; Reyes, E.G.; Roubtsov, V.N.;  On $(t_2,t_3)$-Zakharov–Shabat equations of 
         generalized Kadomtsev-Petviashvili hierarchies. \emph{J. Math. Phys.} {\bf 63}  no. 9, Paper No. 
         093501, 11 pp. (2022)
\bibitem{MRR-jets} Magnot, J-P.; Reyes, E.G., Rubtsov, V.N.; Infinite order structures on differential equations 
        \textit{in preparation}, 2022.
\bibitem{MJD2000} Miwa, T.; Jimbo, M.; Date, E.; \textit{Solitons, Differential Equations, Symmetries 
		and Infinite Dimensional Algebras} (translated by Miles Reid) Cambridge University Press (2000) 
\bibitem{M1} Mulase, M.; Complete integrability of the Kadomtsev-Petvishvili equation.
		{\em Advances in Math.} \textbf{54}  57--66 (1984).
\bibitem{M2} Mulase, M.;  Cohomological structure in soliton equations and
		Jacobian varieties. {\em J. Diff. Geom.} \textbf{19}  403--430 (1984).
\bibitem{M3} Mulase, M.; Solvability of the super KP equation and a generalization of the Birkhoff
		decomposition. {\em Invent. Math.} \textbf{92} 1--46 (1988). 
\bibitem{Nt2005} Ntumba, P.; Sikorski and Fr\"olicher CW-complexes compared. \emph{Demonstratio Math.} 
        {\bf 38 }, no. 1, 207–221 (2005).
\bibitem{Sou} Souriau, J-M.; un algorithme g\'en\'erateur de structures quantiques \textbf{Ast\'erisque} 
	    (hors s\'erie) 341-399 (1985)
\bibitem{St} Stacey, A.; Comparative Smootheology. \emph{Theory and Applications of Categories} {\bf 25}, no.
        4, 64--117 (2011).
\bibitem{Vi1984} Vinogradov, A.M.; The $\mathcal{C}-$spectral sequence, Lagrangian formalism and conservation 
         laws \emph{J. Math. Anal. Appl.} \textbf{100} 1--129 (1984)
\bibitem{Vin1984} Vinogradov, A. M.; Local symmetries and conservation laws. \emph{Acta Appl. Math.} {\bf 2} , 
         no. 1, 21–78 (1984).
\bibitem{Vi} Vinogradov, A.M.; {\em Cohomological Analysis of Partial Differential Equations and Secondary 
			Calculus} . Translations of Mathematical Monographs 204, AMS, Providence (2001).
\bibitem{Vin2013} Vinogradov, A.M.; What are symmetries of PDEs and what are PDEs themselves? Sophus Lie and 
         Felix Klein: the Erlangen program and its impact in mathematics and physics, 137–190, 
         IRMA Lect. Math. Theor. Phys., 23, Eur. Math. Soc., Zürich, (2015).
\bibitem{Wa} Watts, J.; \textit{Diffeologies, differentiable spaces
			and symplectic geometry} PhD thesis, university of Toronto (2012) arXiv:1208.3634

\end{thebibliography}

\end{document}